\documentclass[structabstract]{aa}  
\usepackage{graphicx}%[draft]
\usepackage{natbib}

\bibpunct{(}{)}{;}{a}{}{,}

\usepackage[colorlinks=true,linkcolor=blue,citecolor=blue,urlcolor=black]{hyperref}
\usepackage{txfonts}
\begin{document}

\title{ Revealing the inner circumstellar disk of the T Tauri star S
  CrA N using the VLTI \thanks{Based on observations made with ESO
    telescopes at the La Silla Paranal Observatory under program IDs
    081.C-0272(A), 083.C-0236(C)} }
\titlerunning{The inner cirumstellar disk of the T Tauri star S CrA N}

\author{J. Vural \inst{1} \thanks{Member of the International Max
    Planck Research School (IMPRS) for Astronomy and Astrophysics at
    the Universities of Bonn and Cologne}, A. Kreplin \inst{1} ,
  S. Kraus \inst{2}, G. Weigelt \inst{1}, T. Driebe \inst{3},
  M. Benisty \inst{4}, M. Dugu\'e \inst{5}, F. Massi \inst{6},
  J.-L. Monin \inst{7}, M. Vannier \inst{5}} 
\authorrunning{J. Vural et al.}

\institute{ 
  Max-Planck-Institut f\"ur Radioastronomie, Auf dem H\"ugel 69, 53121
  Bonn, Germany
 \and University of Michigan, Department of Astronomy, 918 Dennison
  Building, 500 Church Street, Ann Arbor, MI 48109-1090, USA
 \and Deutsches Zentrum f\"ur Luft- und Raumfahrt e.V., K\"onigswinterer
  Str. 522-524, 53227 Bonn, Germany
 \and Max-Planck-Institut f\"ur Astronomie, K\"onigstuhl 17, 69117
  Heidelberg, Germany
 \and Laboratoire Lagrange, UMR7293, Universit\'e de Nice
  Sophia-Antipolis, CNRS, Observatoire de la C\^ote d'Azur, 06300 Nice, France 
 \and INAF - Osservatorio Astrofisico di Arcetri, Largo E. Fermi, 5, 
  50125 Firenze, Italy
 \and UJF-Grenoble 1 / CNRS-INSU, Institut de Plan{\'e}tologie et
  d'Astrophysique de Grenoble (IPAG) UMR 5274, Grenoble, F-38041, France
}

\date{Received 26 January 2012 / Accepted 12 June 2012}

\abstract
 {}
{ We investigate the structure of the circumstellar disk of the T
  Tauri star S~CrA~N and test whether the observations agree with the
  standard picture proposed for Herbig Ae stars. }
{ Our observations were carried out with the VLTI/AMBER instrument in
  the {\textit H} and {\textit K} bands with the low spectral
  resolution mode. For the interpretation of our near-infrared AMBER
  and archival mid-infrared MIDI visibilities, we employed both
  geometric and temperature-gradient models. }
{ To characterize the disk size, we first fitted geometric models
  consisting of a stellar point source, a ring-shaped disk, and a halo
  structure to the visibilities. In the {\textit H} and {\textit K}
  bands, we measured ring-fit radii of $0.73 \pm 0.03$~mas
  (corresponding to $0.095 \pm 0.018$~AU for a distance of 130~pc) and
  $0.85 \pm 0.07$~mas ($0.111 \pm 0.026$~AU), respectively. This
  {\textit K}-band radius is approximately two times larger than the
  dust sublimation radius of $\approx$$0.05$~AU expected for a dust
  sublimation temperature of 1500~K and gray dust opacities, but
  approximately agrees with the prediction of models including
  backwarming (namely a radius of $\approx$$0.12$~AU). The derived
  temperature-gradient models suggest that the disk is approximately
  face-on consisting of two disk components with a gap between star
  and disk. The inner disk component has a temperature close to the
  dust sublimation temperature and a quite narrow intensity
  distribution with a radial extension from 0.11~AU to 0.14~AU. }
{ Both our geometric and temperature-gradient models suggest that the
  T~Tauri star S~CrA~N is surrounded by a circumstellar disk that is
  truncated at an inner radius of $\approx$$0.11$~AU. The narrow
  extension of the inner temperature-gradient disk component implies
  that there is a hot inner rim. }

   \maketitle
%
%________________________________________________________________

\section{Introduction} \label{kapint}
Near- and mid-infrared interferometry is able to probe the inner
regions of the circumstellar disks of young stellar objects (YSO) with
unprecedented spatial resolution. However, the detailed structure of
the inner gas and dust disks is not yet well-known. In particular, the
disks of T~Tauri~stars (TTS) are difficult to study because of their
lower apparent brightnesses and the difficulty in spatially resolving
them (e.g. \citealt{2005akebod,2005akewal}). It is not yet known, for
example, whether there is a puffed-up inner rim (PUIR) at the inner
edge of TTS, as observed in several Herbig~Ae/Be disks
\citep{2001natpru,2001duldom,2003muzcal,2005monmil,2005ciekes}. Furthermore,
observations suggest that several TTS are surrounded by an additional
extended halo of scattered starlight, which influences the precise
determination of the disk size \citep{2008pinmen}. The positions of
observed TTS in the size-luminosity relation \citep{2007eishil}
suggest, that TTS have slightly larger inner disk radii than
expected. However, if one compares the TTS radii with predictions of
models including backwarming \citep{2007milmal,2010dulmon}, the
discrepancy disappears.

In this paper, we investigate the circumstellar disk of the TTS
\object{S~CrA~N}, which is the more massive star in the binary
S~CrA. The binary separation is approximately 1.4\arcsec
($\approx$150~AU) \citep{1993reizin,1997ghemcc} and its position angle
(PA) is $157\degr$ \citep{1997ghemcc}. The binary components are
coeval and have an age of $\approx$3~Myr \citep{2003pragre}. The
properties of both stars are listed in Table~\ref{tabpro}.

S~CrA~N is a classical TTS \citep{2006mccghe}. Its infrared excess
suggests the presence of a dusty disk. The precise determination of
its spectral type is difficult owing to a strong veiling of the
absorption lines \citep{1961bon}. \citet{2006mccghe} inferred a
spectral type of K3 and \citet{2007carvan} obtained a spectral type of
G5Ve.  The veiling as well as the detection of a strong Br$\gamma$
flux suggest the presence of an accretion disk
\citep{2003pragre,2006mccghe}.  \citet{2009schwol} resolved the disk
of S~CrA~N in the mid-infrared with VLTI/MIDI and modeled the spectral
energy distribution (SED) and visibilities with the Monte Carlo code
MC3D to constrain several disk parameters.

The S~CrA system is probably connected with Herbig-Haro objects:
HH~82A and B are oriented towards a position angle of
$\approx$95$\degr$, whereas HH~729A, B and C lie in the direction of
$\approx$115$\degr$ \citep{1988reigra}. The different PA can, for
example, be explained by either two independent outflows from each of
the binary components of S~CrA or by regarding the HH objects as the
edges of an outflow cavity \citep{2004wanmun}. The determination of
the orientation of the circumstellar disk might clarify the exact
relationship.

\citet{2005walmin} found that the secondary, S~CrA~S, can be brighter
in the optical than the primary for up to one third of the time and
that S~CrA has one of the most rapidly varying brightnesses of the
TTS. This variability is discussed in \citet{1992gra}, who proposes
that it is caused by geometrical obscuration as well as accretion
processes and emphasizes the probability of clumpy
accretion.

Only very recently, \citet{2010ortsug} observed light echoes in the
reflection nebula around S~CrA and suggested that the structure is
similar to the Oort cloud in our solar system. With their method, they
also determined the distance of S~CrA (138$\pm$16~pc) confirming the
previously derived distance of 130~pc \citep{2003pragre,2008neufor}.

\begin{table}[t!]
\caption{Properties of the S CrA binary components}     
\label{tabpro}  
\begin{minipage}{0.5\textwidth} 
\centering                       
\begin{tabular}{l|r|r} 
\hline\hline
Parameter & S CrA N & S CrA S \\ \hline
spectral type & K3 & M0 \\
M$_*$ [M$_\odot$] & $1.5\pm 0.2$ & $0.6\pm 0.2$  \\
T$_*$ [K] & $4800\pm 400$ & $3800\pm 400$  \\
L$_*$ [L$_\odot$] & $2.30\pm 0.70$ & $0.76\pm 0.24$  \\
Br$\gamma$ [$10^{-16}$ Wm$^{-2}$]&$ 4.23\pm 1.10$ & $1.59\pm 0.57$ \\
m$_\mathrm J$ [mag] & 8.6 & 9.4 \\
m$_\mathrm H$ [mag] & 7.5 & 8.3 \\
m$_\mathrm K$ [mag] & 6.6 & 7.3 \\
distance [pc]& \multicolumn{2}{c}{$130\pm 20$} \\
A$_\mathrm v$ &  \multicolumn{2}{c}{2.8\tablefootmark{a}} \\
binary sep. [\arcsec]& \multicolumn{2}{c}{$1.30\pm0.05$\tablefootmark{b}/1.4 \tablefootmark{c} }\\\hline
\end{tabular}
\tablefoot{If not mentioned otherwise, the values are taken from
  \citet{2003pragre}. For the spectral type of the primary,
  \citet{1988herbel} found K6 and \citet{2007carvan}
  G5Ve. \citet{2010ortsug} found a distance of 138$\pm$16~pc, which is
  consistent with the table value and the one extensively discussed by
  \citet{2008neufor} (130~pc).\\ Other references:
  \tablefoottext{a}{\citet{1998pat}},
  \tablefoottext{b}{\citet{2006mccghe}},
  \tablefoottext{c}{\citet{1997ghemcc}} }
\end{minipage}
\end{table}

In this paper, we present VLTI/AMBER observations and the
temperature-gradient modeling of the circumstellar disk of S~CrA~N. In
Sect.~\ref{kapodr}, we describe the observations and data
reduction. In Sect.~\ref{kapmod}, we describe our modeling, which is
discussed in Sect.~\ref{kapdis}.

%__________________________________________________________________

\section{Observation and data reduction}\label{kapodr}
\begin{figure}[tb]
  \centering
  \includegraphics[angle=-90,width=0.5\textwidth]{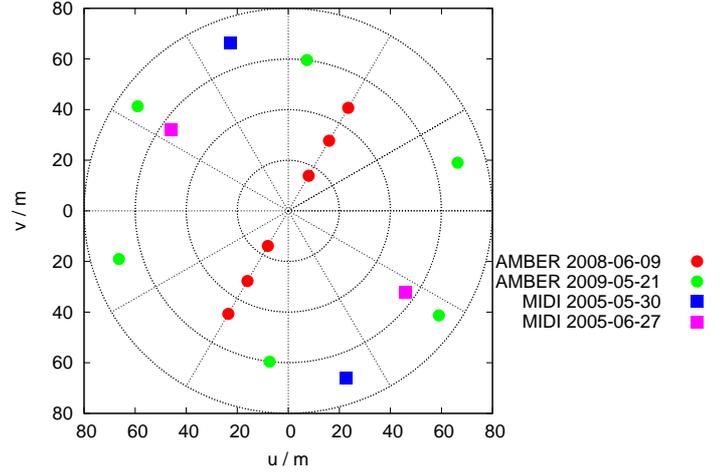}
  \caption{The uv coverage of the AMBER (red and green dots) and MIDI
    (blue and pink squares) observations used for modeling.}
  \label{figuvc}
\end{figure}

Our observations (program IDs 081.C-0272(A) and 083.C-0236(C)) were
carried out with the near-infrared interferometry instrument AMBER
\citep{2007petmal} of the Very Large Telescope Interferometer (VLTI)
in the low-spectral-resolution mode ($R=30$). The observational
parameters of our data sets are listed in Table~\ref{tabobs2} and the
uv coverage is shown in Fig.~\ref{figuvc}. All data were recorded
without using the FINITO fringe tracker. The observations were
performed with projected baselines in the range from 16~m to 72~m.

For data reduction, we used \textit{amdlib}-3.0
\footnote{\url{http://www.jmmc.fr/data_processing_amber.htm}}.  A
fraction of the interferograms were of low quality. We therefore
selected the 20\% data with the highest fringe signal-to-noise-ratio
of both the target and the calibrator interferograms to improve the
visibility calibration \citep{2007tatmil}. Furthermore, it was
impossible to reduce the {\it H} band data of data set I. We applied
an equalisation of the optical path difference histograms of
calibrator and target to improve the visibility calibration
\citep{2012krekra}.

\begin{figure}[t!]
  \centering
  \includegraphics[angle=-90,width=0.4\textwidth]{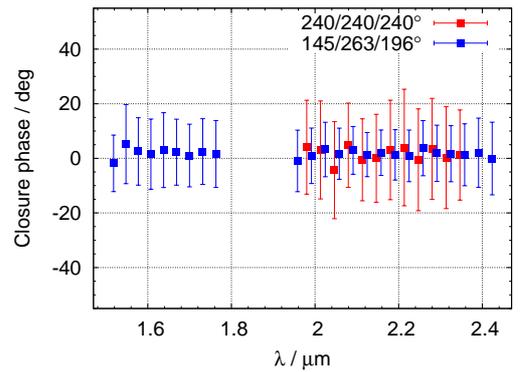} 
  \caption{Closure phases measured with AMBER in the {\it H} and {\it
      K} bands.}
  \label{figcp}
\end{figure}

The derived visibilities are shown in Figs.~\ref{figdia} and
\ref{figtgm2}. The measured closure phase (Fig.~\ref{figcp}) is
approximately zero, which suggests that it is an approximately
symmetric object. However, small closure phases are also expected
because the object is only partially resolved.

\begin{table*}[hbt]
\caption{AMBER observations}
\label{tabobs2}   
\centering                       
\begin{tabular}{cccccccll}  
 \hline \hline
Data set & Night & B$_\mathrm{proj}$ [m] & PA [\degr] & Seeing [\arcsec] & Airmass & DIT [ms] & Calibrator & Calibrator diameter [mas] 
\\ \hline
I & 2008-06-09 & 16/32/47 & 240/240/240 & 0.91 & 1.07 & 100 & HD 183\,925 & 1.44$\pm$0.02 \tablefootmark{a} \\ 
II & 2009-05-21 & 60/69/72 & 145/263/196 & 0.98 & 1.06 & 150 & HD 170\,773 & 0.38$\pm$0.026 \tablefootmark{b} \\ \hline
\end{tabular}
\tablefoot{The parameters of the MIDI observations used for the
  modeling in Sect.~\ref{kaptgm} are described in \citet{2009schwol}
  and the uv coverage is shown in Fig.~\ref{figuvc}. References:
  \tablefoottext{a}{\citet{2005ricper}},
  \tablefoottext{b}{\citet{2009kharoe}}}
\end{table*}
%______________________________________________________________

\section{Modeling} \label{kapmod}
%%%%%%%%%%%%%%%%%%%%%%%%%%%%%%%%%%%%%%%%%%%%%%%%%%%%%%%%%%%%%%%%%%%%%
\subsection{Geometric models}\label{kapgeo}

\begin{figure*}[tb]
  \centering
  \begin{minipage}{0.4\textwidth}
  \includegraphics[angle=-90,width=1\textwidth]{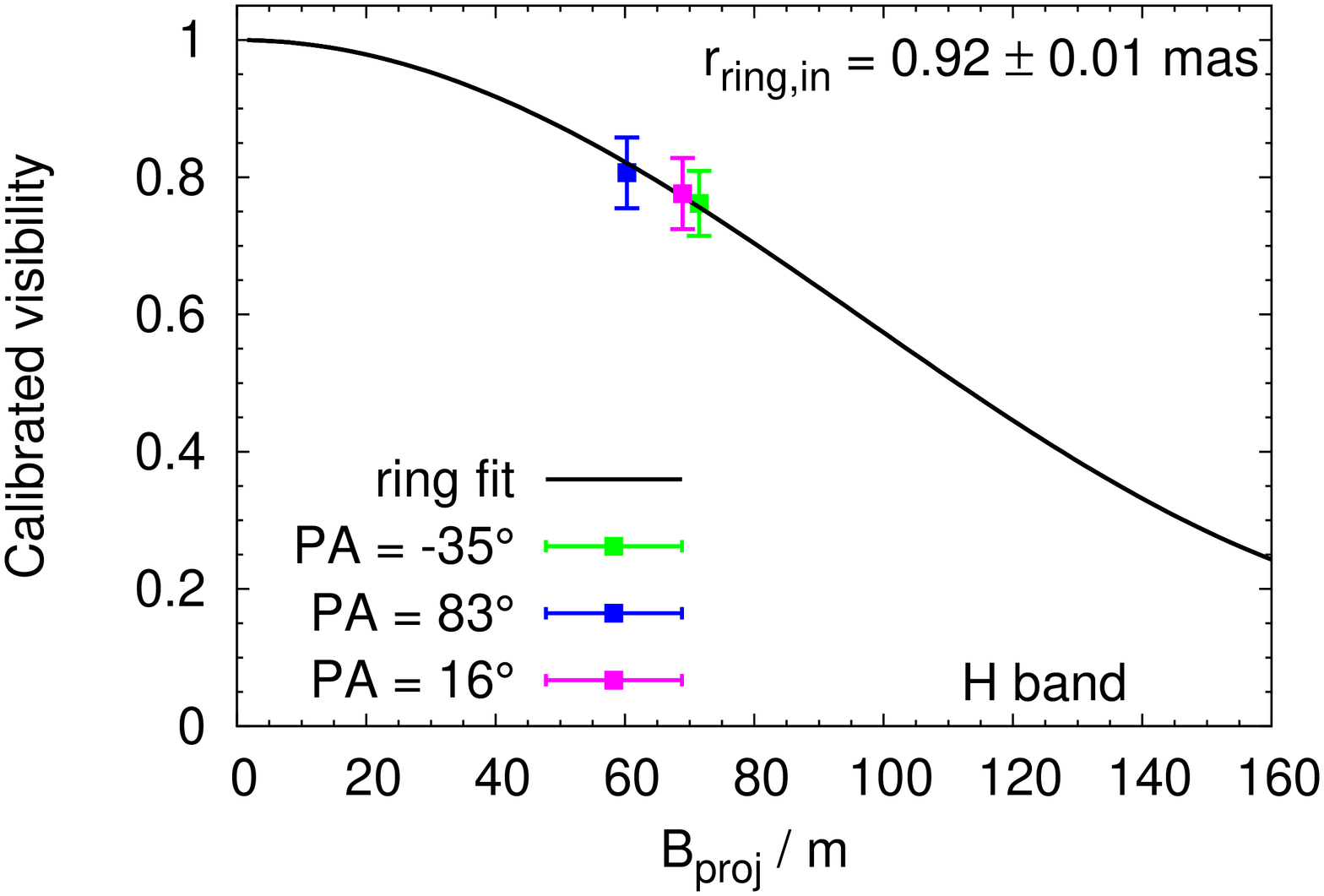}
  \includegraphics[angle=-90,width=1\textwidth]{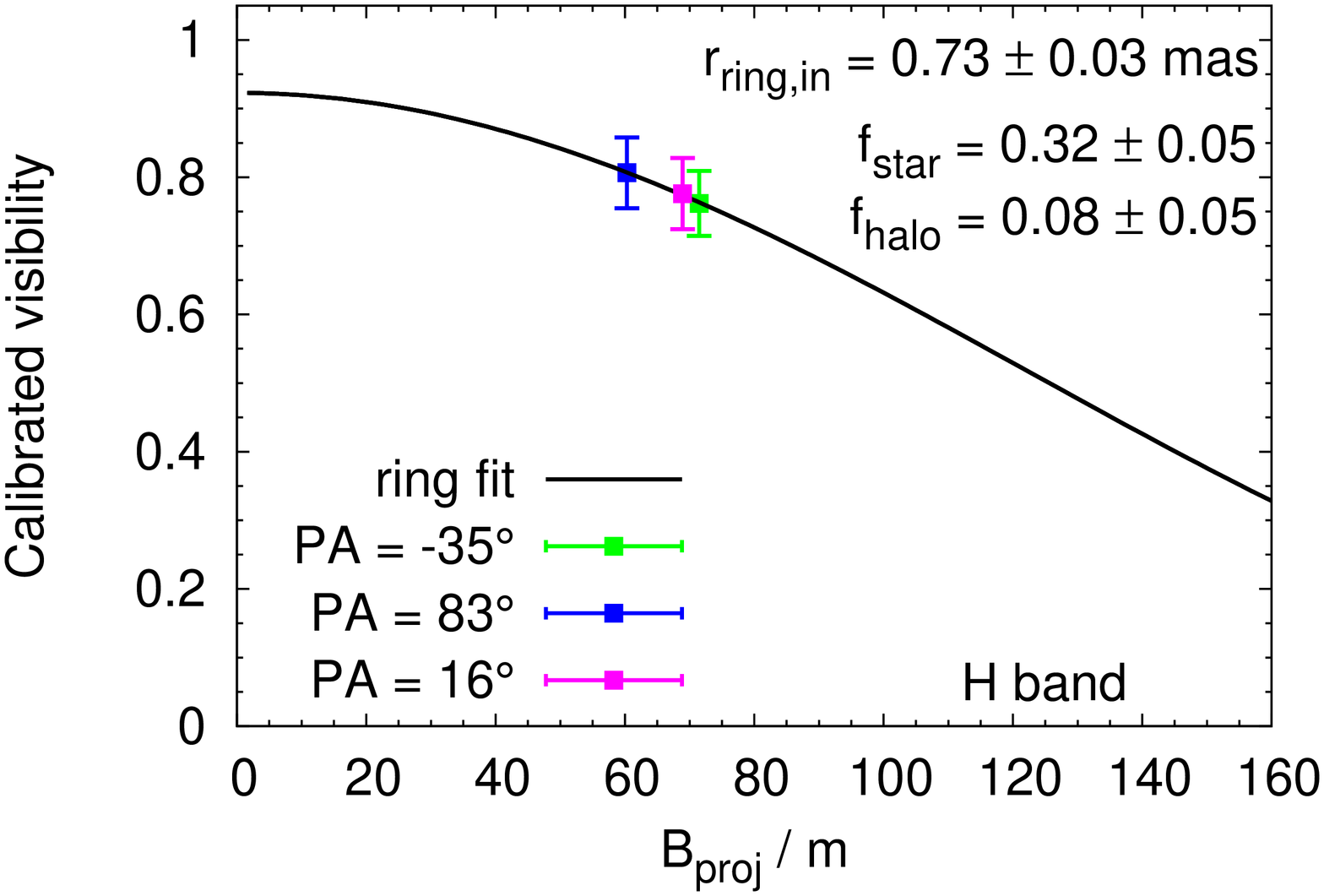}
  \end{minipage}
  \begin{minipage}{0.4\textwidth}
  \includegraphics[angle=-90,width=1\textwidth]{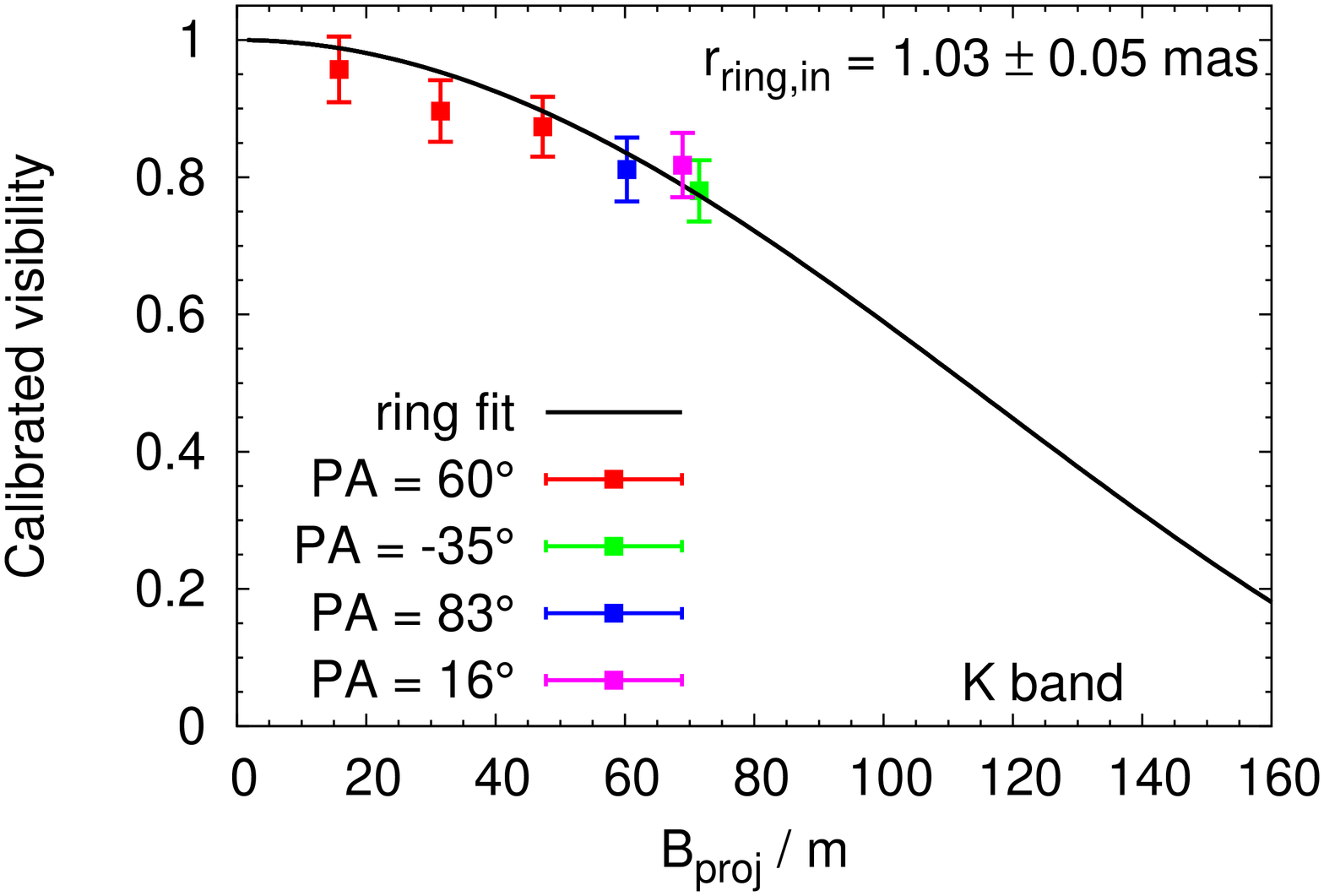}
  \includegraphics[angle=-90,width=1\textwidth]{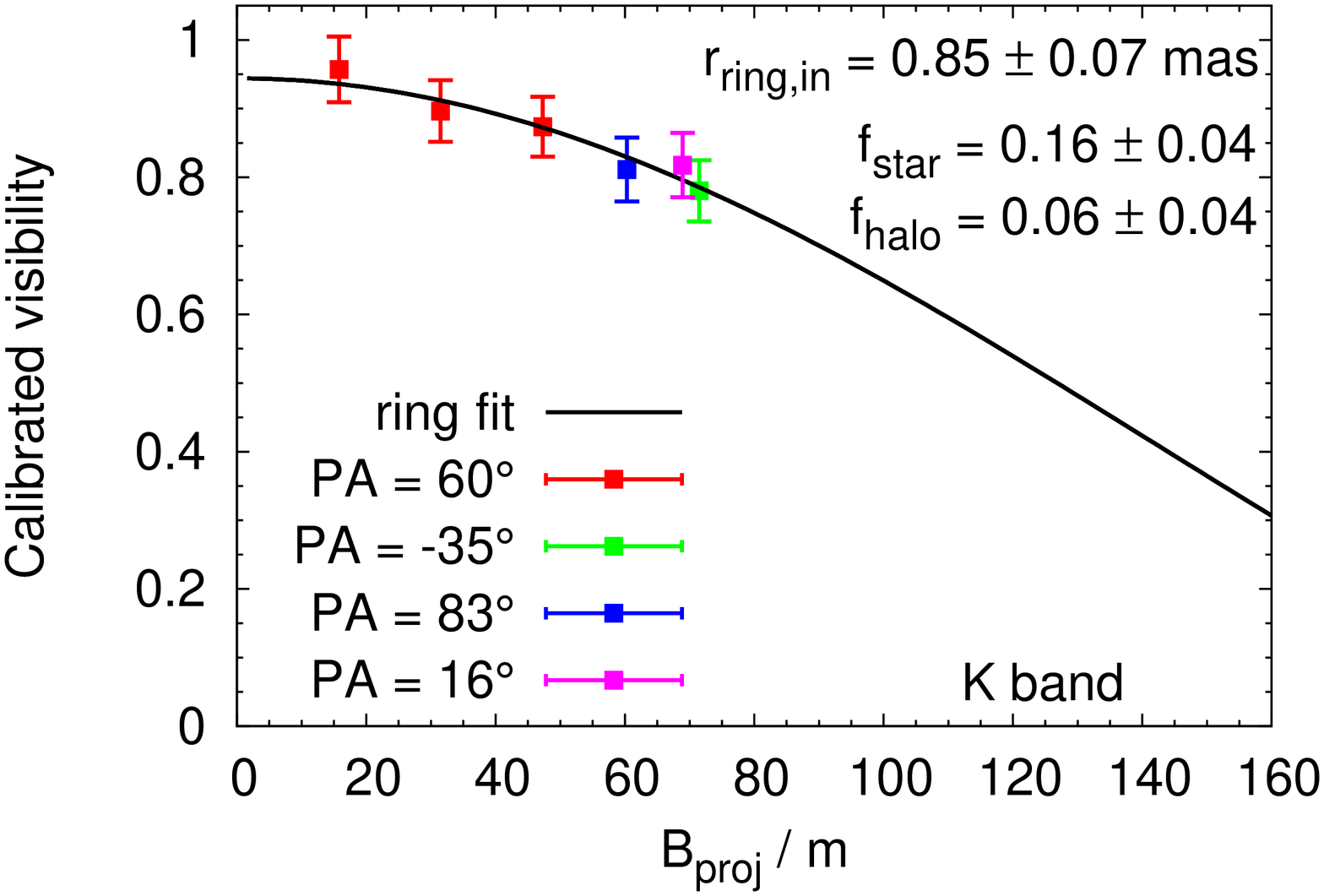}
  \end{minipage} 
  \caption{{\it Top:} {\it H}-band (left) and {\it K}-band (right)
    visibilities of S~CrA~N together with ring-fit models (described
    in Sect.~\ref{kapgeo}) consisting of a circular symmetric ring
    (ring width = 20\% of inner ring radius {$r_\mathrm{ring,in}$}) and an unresolved stellar
    source. {\it Bottom:} Models consisting of the same star+ring
    model as above, plus a fully resolved halo component.}
  \label{figdia}
\end{figure*}

To measure the characteristic size of the circumstellar environment in
the {\it H}- and {\it K}-bands, we fitted geometric models to the
visibilities. Figure~\ref{figdia} shows the observed visibilities and
the fitted models, which consist of a ring (the ring width being 20\% of
the inner radius; \citealt{2003eislan,2005monmil}), a stellar point
source, and an extended fully resolved halo (in only the lower two
panels). This extended halo is assumed to represent the stellar light
scattered off a large-scale circumstellar structure
\citep{2005akewal}. The motivation for a ring fit is that the
model intensity distribution is expected to have a dominant PUIR
brightness \citep{2001natpru,2001duldom}. Ring-fit radii are often
used in the literature to characterize the disk size of YSO and to
discuss their location in the size-luminosity relation
(e.g. \citealt{2002monmil,2010dulmon}).

The flux contribution $f_\mathrm{star}+f_\mathrm{halo}$ from the star
($f_\mathrm{star}$) plus a halo of scattered starlight
($f_\mathrm{halo}$) was derived from the SED fit in Fig.~\ref{figtgm2}
and amounts to 0.22 of the total flux
(i.e. $f_\mathrm{star}+f_\mathrm{halo}+f_\mathrm{disk}$) in the {\it
  K} band (2.2~$\mu$m) and 0.40 in the {\it H} band (1.6~$\mu$m). The
total visibility $V$ can be described by
\begin{equation}
|V| = |(1-f_\mathrm{star}-f_\mathrm{halo})V_\mathrm{disk}
  + f_\mathrm{star}V_\mathrm{star}
  + f_\mathrm{halo}V_\mathrm{halo}| \;,
\end{equation}
where $V_\mathrm{star}=1$ and $V_\mathrm{halo}=0$.

In fitting the data, we averaged the visibility data in each of the two bands.
The {\it H} band visibilities were averaged over the wavelength range of
$\approx$1.55--1.75~$\mu$m and the {\it K} band visibilities over
$\approx$1.95--2.40~$\mu$m.

For the inner fit radius $r_\mathrm{ring,in}$, we obtained $0.73 \pm
0.03$~mas in the {\it H} band and $0.85 \pm 0.07$~mas in the {\it K}
band. The linear radii in AU for a distance of 130~pc (see
Table~\ref{tabpro}) are listed in Table~\ref{tabrad}. 

\begin{table*}[thb]
  \centering  
  \caption{Ring-fit parameters in the {\it H} and {\it K} bands. The
    error in the distance measurement ($130\pm 20$~pc, see
    Table~\ref{tabpro}) is included in the errors in the radii.}
  \label{tabrad}                     
  \begin{tabular}{lccccc}  
    \hline \hline
    Model          & Band    & $r_\mathrm{ring,in}$ & $f_\mathrm{halo}$& $f_*$ & $f_\mathrm{disk}$  \\ 
                   &         &  [AU]               \\ \hline
    ring-star      & {\it H} & $0.120 \pm 0.020$  & 0                        & $0.40\pm 0.09$  & $0.60\pm 0.09$ \\
    ring-star      & {\it K} & $0.134 \pm 0.027$  & 0                        & $0.22\pm 0.05$  & $0.78\pm 0.05$ \\
    ring-star-halo & {\it H} & $0.095 \pm 0.018$  & $0.08 \pm 0.05$ & $0.32 \pm 0.14$ & $0.60\pm 0.09$ \\
    ring-star-halo & {\it K} & $0.111 \pm 0.026$  & $0.06 \pm 0.04$ & $0.16 \pm 0.09$ & $0.78\pm 0.05$ \\
 \hline
  \end{tabular}
\end{table*}

%%%%%%%%%%%%%%%%%%%%%%%%%%%%%%%%%%%%%%%%%%%%%%%%%%%%%%%%%%%%%%%%%%%%%%%%%

To test whether an elongation of the object can be measured, we also
fitted an elliptic ring model. However, we were unable to detect any
significant elongation. Because of the errors in the data, we can only
derive a rough estimate of the elongation of $<30\%$. Therefore, we
assume in the following modeling that the disk is oriented
approximately face-on.

%%%%%%%%%%%%%%%%%%%%%%%%%%%%%%%%%%%%%%%%%%%%%%%%%%%%%%%%%%%%%%%%%%%%%
\subsection{Temperature-gradient disk model} \label{kaptgm}

\begin{table*}[th]
  \centering  
  \caption{Overview of the parameter space scanned for the
    temperature-gradient models and parameters (with errors) of the
    best-fit models for one temperature-gradient disk (A, B) and two
    temperature-gradient disk components (C, D, E). For details about
    the individual models, we refer the reader to Sect.~\ref{kaptgm}.}
  \label{tabpar}                     
  \begin{tabular}{lrrrrrr}  
    \hline \hline
    Parameter        & Scan range        & A & B & C & D & E  \\\hline
    Constraints      &                   &   &   & $r_\mathrm{out,1} = r_\mathrm{in,2}$ & $r_\mathrm{out,1} = r_\mathrm{in,2}$ &   \\ 
                     &                   &   &   & $T_\mathrm{out,1} = T_\mathrm{in,2}$ & & \\\hline
    $r_\mathrm{in,1}$ [AU]  & 0.01--0.3  & ...                            & ...                         &$0.11^{+0.06}_{-0.04}$ &$0.14^{+0.03}_{-0.07}$ &$0.13^{+0.04}_{-0.05}$ \smallskip \\
    $r_\mathrm{out,1}$ [AU] & 0.02--4    & ...                            & ...                         &$0.14^{+0.05}_{-0.03}$ &$0.17^{+0.09}_{-0.05}$ &$0.17^{+0.09}_{-0.05}$ \smallskip \\ 
    $r_\mathrm{in,2}$ [AU]  & 0.01--0.6  & $0.05\pm0.01$         &$0.05\pm0.03$       &$0.14^{+0.05}_{-0.03}$ &$0.17^{+0.09}_{-0.05}$ &$0.14^{+0.04}_{-0.05}$ \smallskip \\ 
    $r_\mathrm{out,2}$ [AU] &    1--50   & $\ge4$                &$\ge5$              &$\ge26$                &$\ge26$                &$\ge26$                \smallskip \\
    $T_\mathrm{in,1}$ [T]   & 100--3000  & ...                            & ...                         &$1690^{+170}_{-410}$   &$1530^{+330}_{-250}$   &$1500^{+360}_{-220}$ \smallskip \\
    $T_\mathrm{in,2}$ [T]   & 200--3000  & $1900^{+90}_{-150}$   &$1890^{+20}_{-140}$ &$660^{+90}_{-160}$     &$633^{+130}_{-120}$    &$660^{+90}_{-60}$    \smallskip \\
    $q_1$                   & 0.35--0.85 & ...                            & ...                         &$0.2\pm0.2$            &$0.5\pm0.4$            &$0.5\pm0.4$  \smallskip \\
    $q_2$                   & 0.3--0.9   & $0.75^{+0.1}_{-0.08}$ &$0.75\pm0.1$        &$0.5\pm0.1$            &$0.5^{+0.12}_{-0.04}$  &$0.5^{+0.12}_{-0.04}$  \smallskip \\
    $f_\mathrm{halo}$       & 0--1       & ...                            &$0.2\pm0.1 $        & ...                            & ...                            & ...  \\
    \hline
    $\chi_\mathrm{red}^2$   &            & 16.9 & 11.5 & 3.2     & 3.0  & 3.0 \\ \hline
  \end{tabular}
\end{table*}

In this section, we present temperature-gradient models to help us
interpret our observations. Each temperature-gradient model is the sum
of many rings that emit blackbody radiation with temperatures
$T(r)$. For the temperature distribution, a power law is assumed
\citep{1992hilstr}, of $T(r)=T_0 \left(r/r_0\right)^{-q}$.  Here,
$T_0$ is the effective temperature at a reference radius $r_0$. The
parameter $q$ depends on the disk morphology and is predicted to be
0.50 for flared irradiated disks and either 0.75 for standard viscous
disks or flat irradiated disks
\citep{1997chigol}. Wavelength-dependent visibilities and fluxes of
these model disks depend on the disk inclination, the inner radius
$r_\mathrm{in}$, the outer radius $r_\mathrm{out}$, the temperature
$T_\mathrm{in}$ at $r_\mathrm{in}$, and the temperature power-law
index $q$. As explained in Sect.~\ref{kapgeo}, we assumed that the
disk is viewed face-on.

To determine the best temperature-gradient model, we calculated the
model disks (one- and two-component structures) for all combinations
of the parameters with scan ranges described in Table~\ref{tabpar}
($\approx$400\,000 models). The total $\chi_\mathrm{red}^2$ is a sum
over the $\chi^2_\mathrm{red}$ of all visibility points for all six
AMBER baselines and two MIDI baselines (see \citealt{2009schwol}), and
the SED. We computed all combinations by running the model for six to
ten steps per parameter and then chose the areas with the smallest
value of $\chi_\mathrm{red}^2$ to obtain a finer mesh. We took into
account only wavelengths with $\lambda < 20\mu$m. The errors in the
best-fit model parameters are 1-$\sigma$~errors.

The model assumes a stellar point source with the parameters
$T_*=$~4800~K, $L_*= 2.3 L_\odot$, distance = 130~pc, and $A_\mathrm
v= 2.8$ (see references for the stellar parameters in
Table~\ref{tabpro}). The model parameters of all derived models (disk
plus star, disk plus star plus halo, as well as several two-component
disks) are listed in Table~\ref{tabpar}.

\subsubsection{One-component temperature-gradient disk models}
The simplest temperature-gradient model~A (see Table~\ref{tabpar})
consists of only one single disk component plus the star. It cannot
reproduce the SED as well as the MIR and NIR visibilities
simultaneously ($\chi^2_\mathrm{red}\approx$$17$).

For the star-disk-halo model B, we added a fully resolved halo ($V=0$)
to the star-disk model A. This halo intensity distribution is assumed
to represent the stellar light scattered off the large-scale
circumstellar material and to have approximately the same SED as the
star itself (cf. \citealt{2005akewal}). Nevertheless, the improvement
achieved was very small since the $\chi^2_\mathrm{red}$ became only
11.5 (see Table~\ref{tabpar}). We also computed 20 star-disk-halo
models with resolved halos of different widths in the range from
0.05~AU to 100~AU, but the $\chi^2_\mathrm{red}$ did not significantly
improve since it remained larger than $\approx$$11.5$ in all cases. As
in the case of model~A without a halo, it was only possible to fit
either the SED and the AMBER measurements or the SED and the MIDI
measurements. This suggests that a more complicated disk structure
should be considered. Therefore, we introduced a second disk component
in the following, but -- as we preferred to adopt a two-disk modeling
with a minimum number of free parameters -- we omitted the weak halo.

\subsubsection{Two-component temperature-gradient disk models}\label{tabtg2}
Model~C is the result of our attempt to find a two-disk component
model with the least number of parameters (Table~\ref{tabpar}). We
therefore introduced constraints on the inner and outer radii and
temperatures of disks 1 and 2, of $T_\mathrm{out,1} = T_\mathrm{in,2}$
and $r_\mathrm{out,1} = r_\mathrm{in,2}$. We obtained
$\chi_\mathrm{red}^2=3.2$ (the power-law index $q_1$ is determined
from the temperature slope between $r_\mathrm{in,1}$ and
$r_\mathrm{out,1}$ and is therefore no longer a free parameter).

We also tested models with more parameters: In model~D,
$r_\mathrm{out,1} = r_\mathrm{in,2}$ is the only constraint and
$\chi_\mathrm{red}^2=3.0$. For model~E, there are no constraints, but
we still get a similar $\chi_\mathrm{red}^2$ (3.0).
Table~\ref{tabpar} shows that only models C to E have a
$\chi_\mathrm{red}^2$ from 3.0 to 3.2 and among those, model~C has the
advantage that it has the smallest number of parameters.

\begin{figure*}[htbp]
  \centering
  \includegraphics[width=0.33\textwidth]{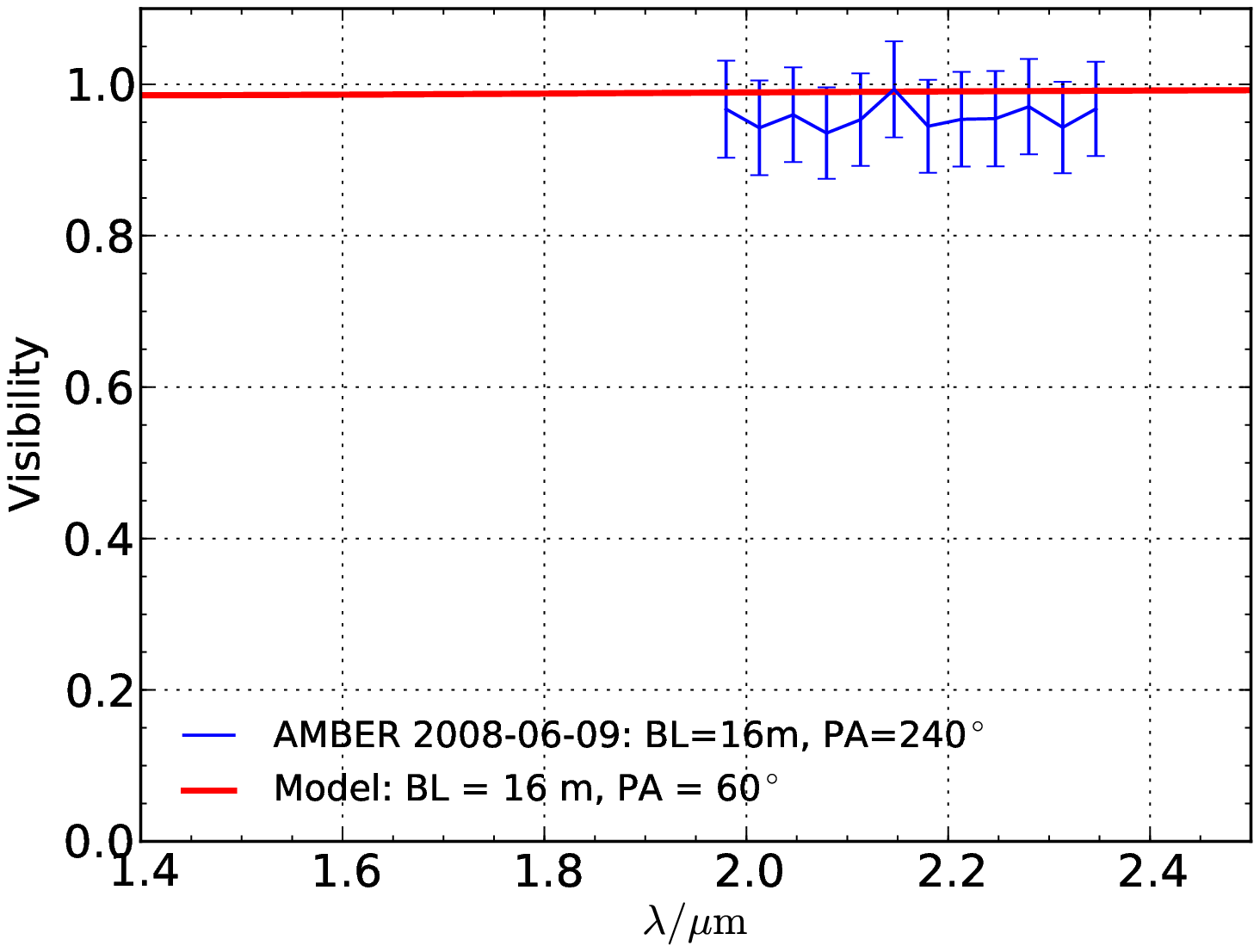}
  \includegraphics[width=0.33\textwidth]{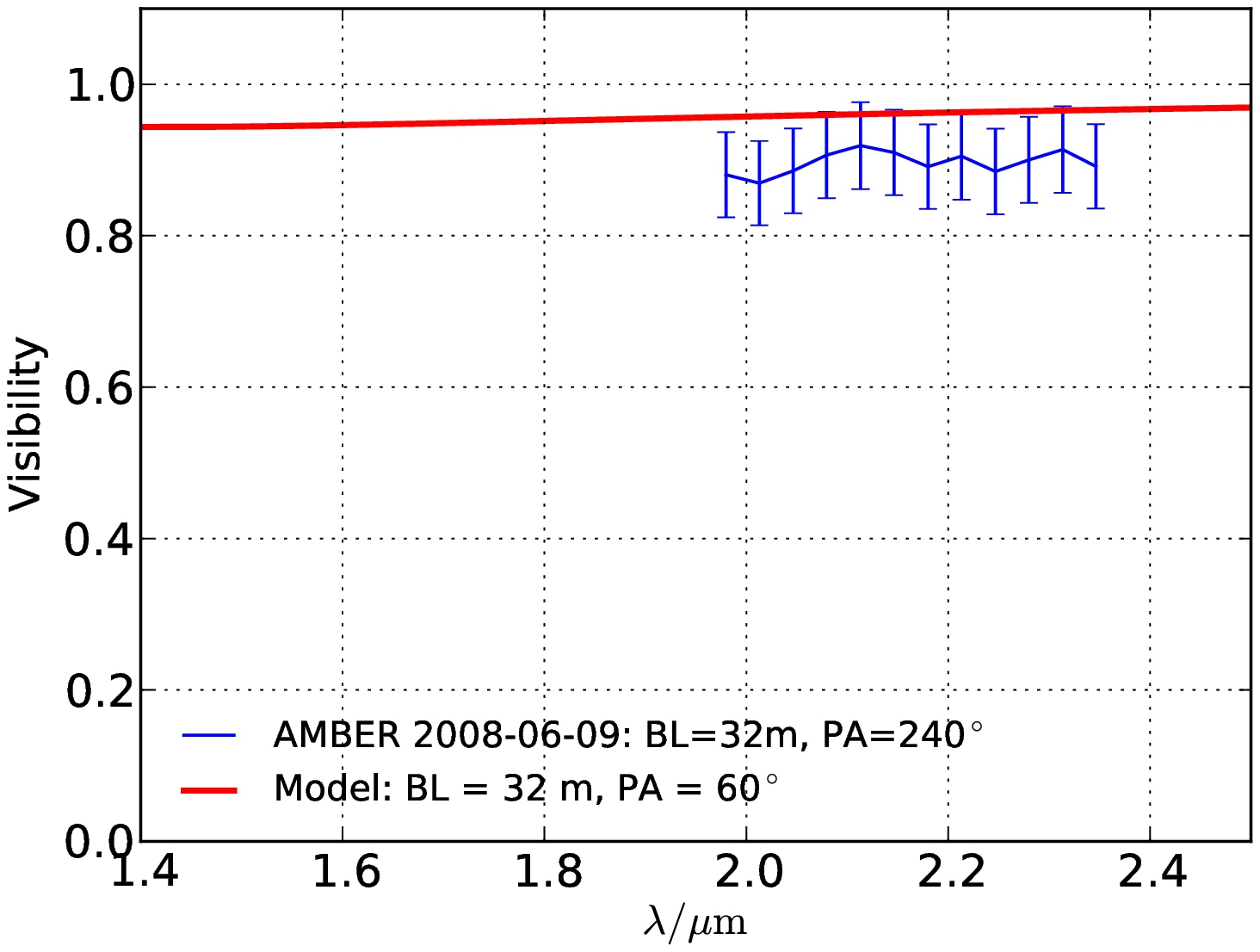}
  \includegraphics[width=0.33\textwidth]{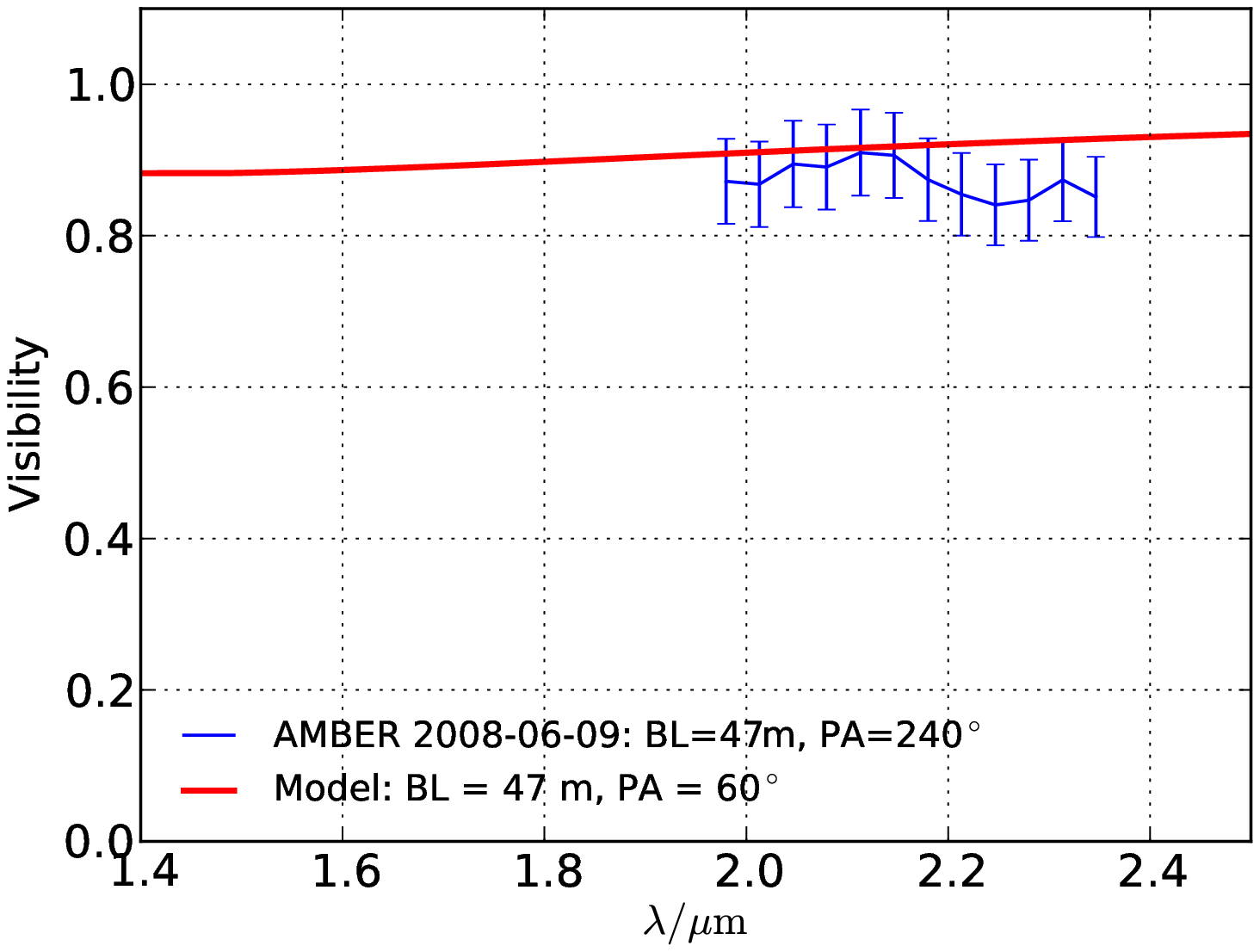}
  \includegraphics[width=0.33\textwidth]{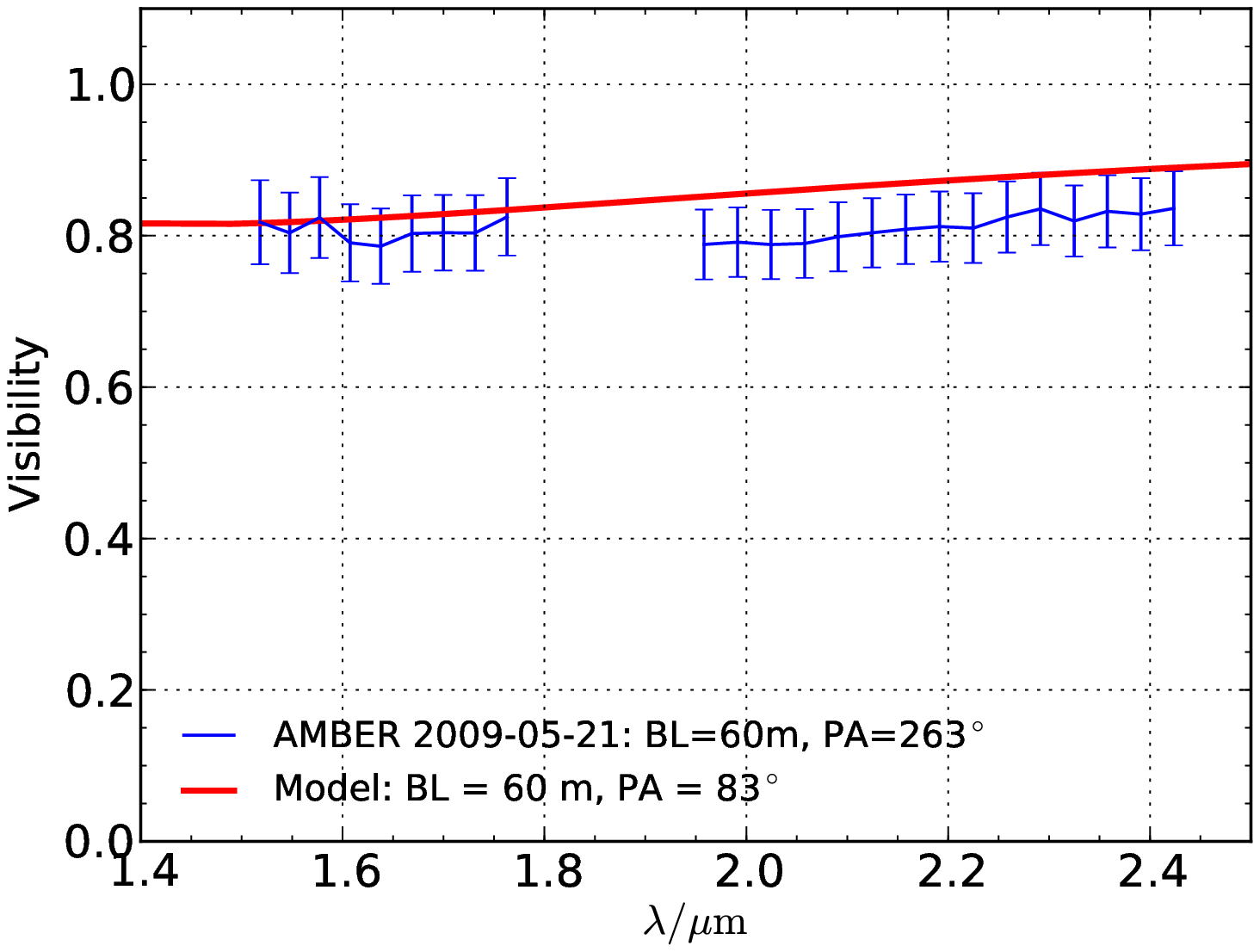}
  \includegraphics[width=0.33\textwidth]{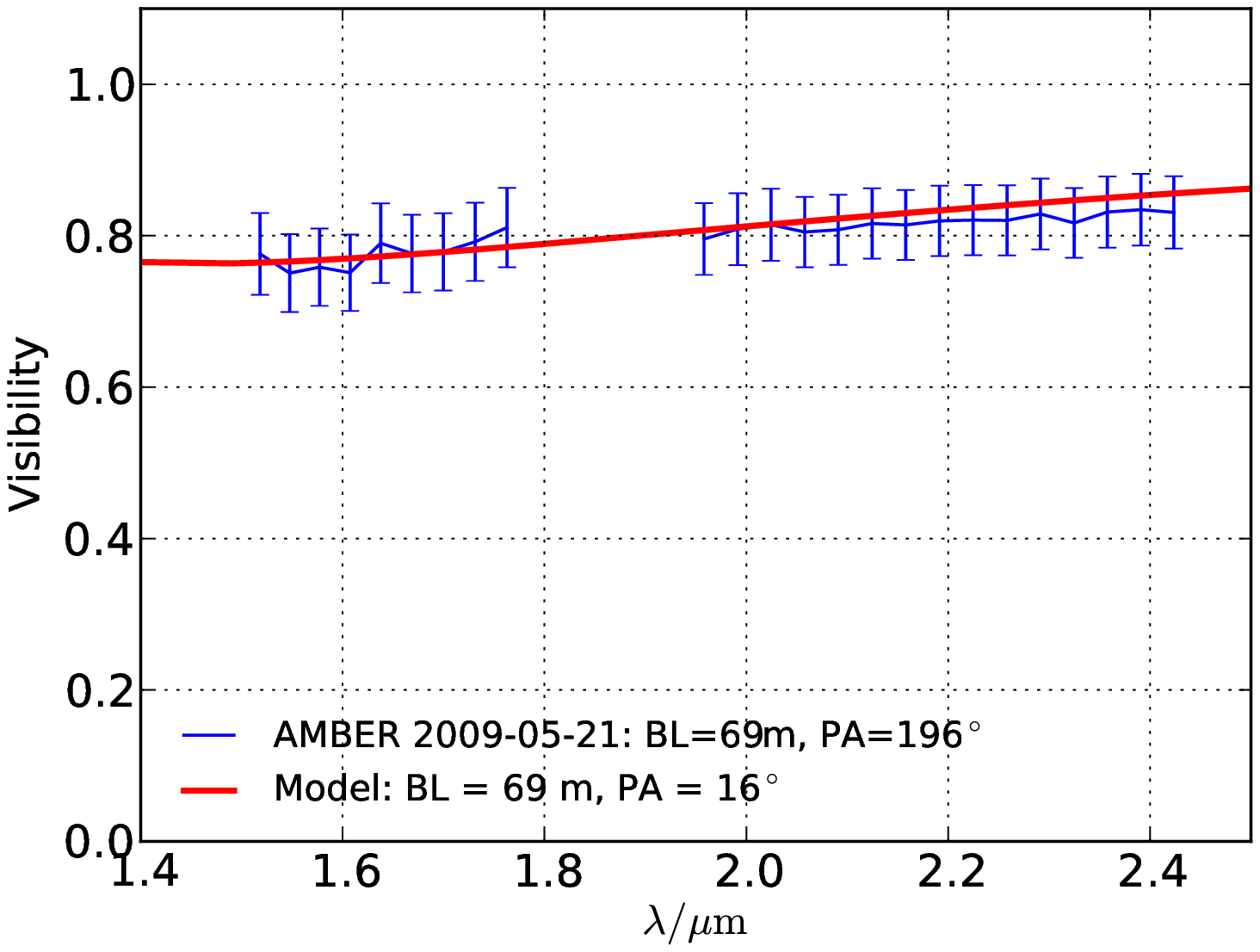}
  \includegraphics[width=0.33\textwidth]{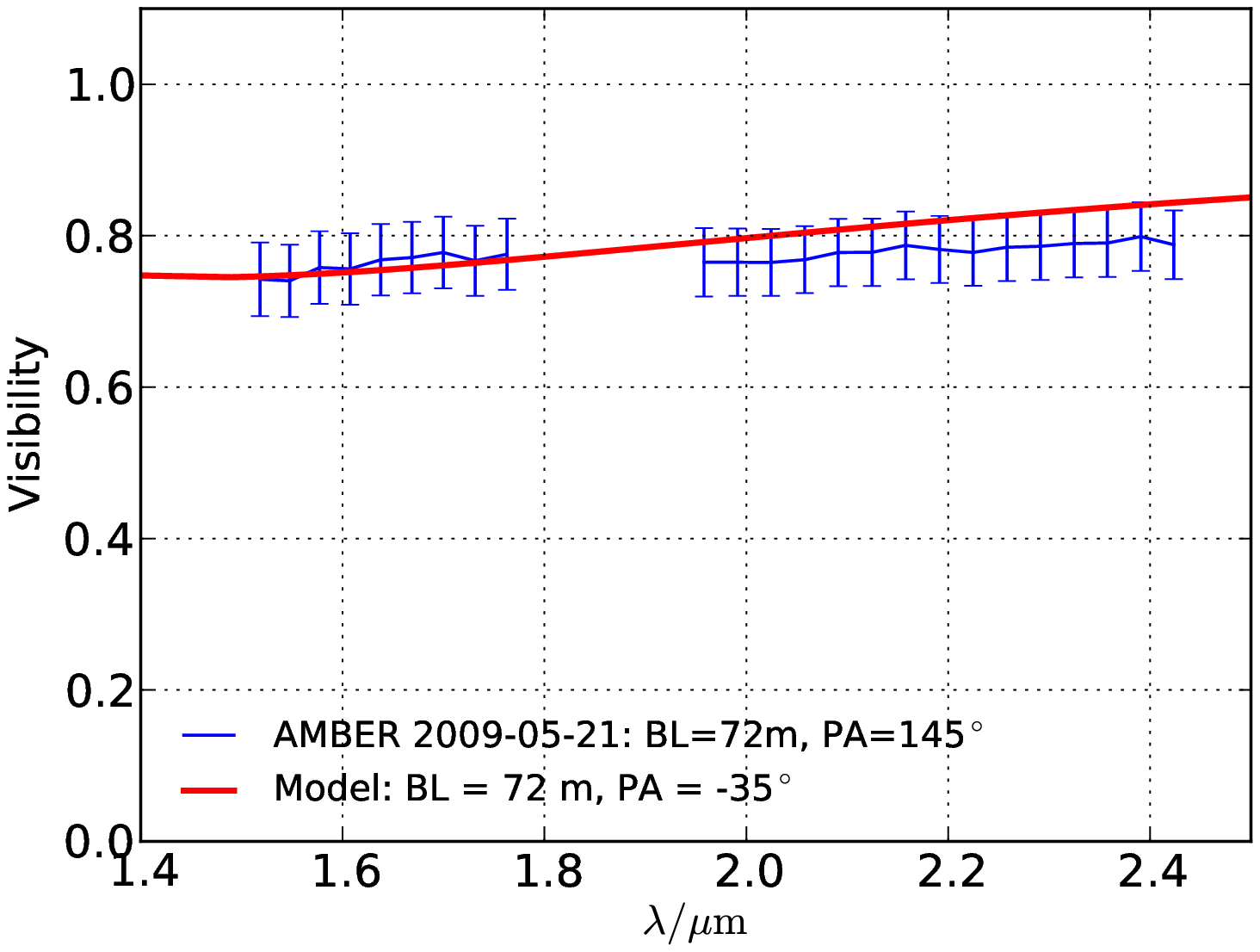}
\vspace*{3mm}
  \includegraphics[width=0.33\textwidth]{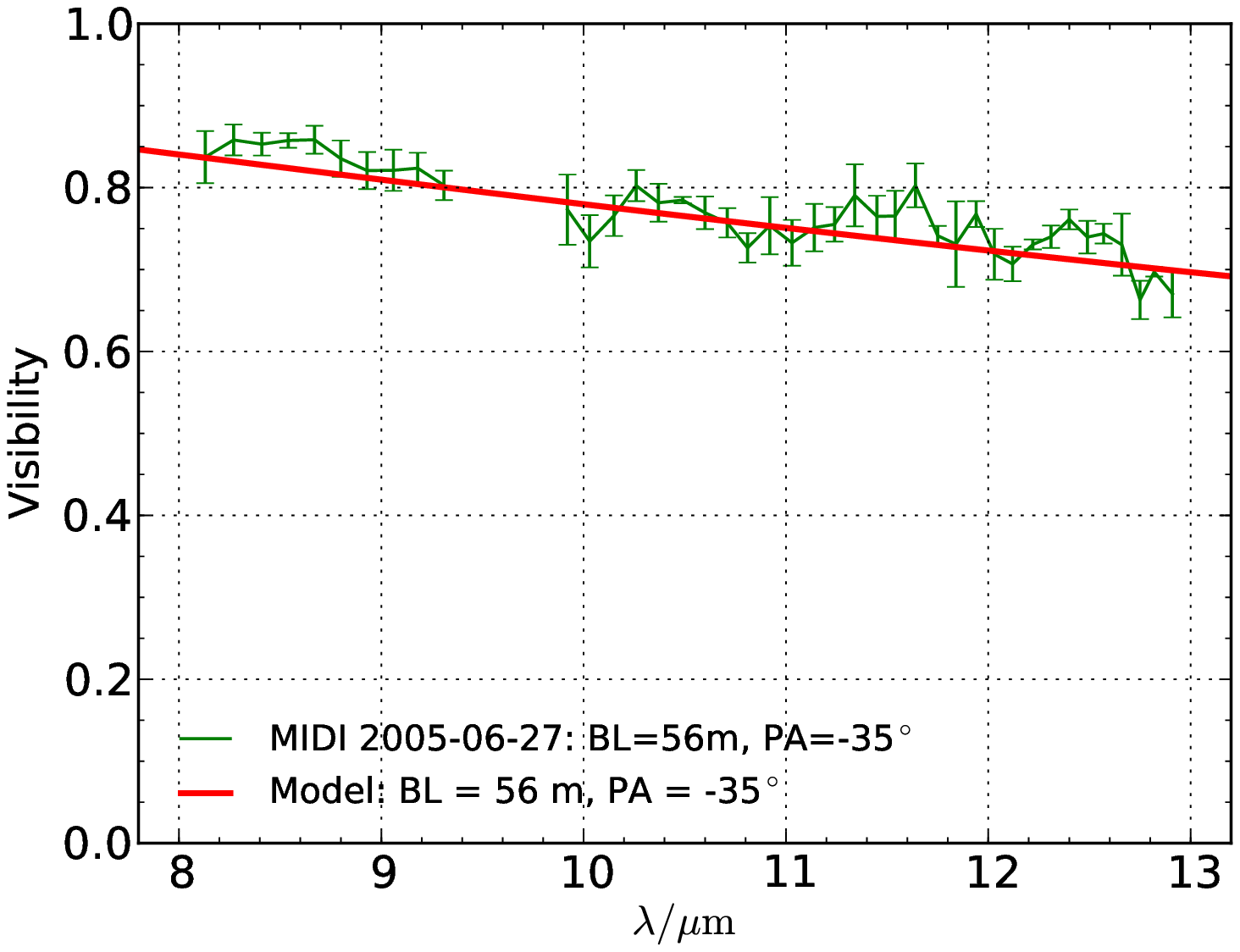}
  \includegraphics[width=0.33\textwidth]{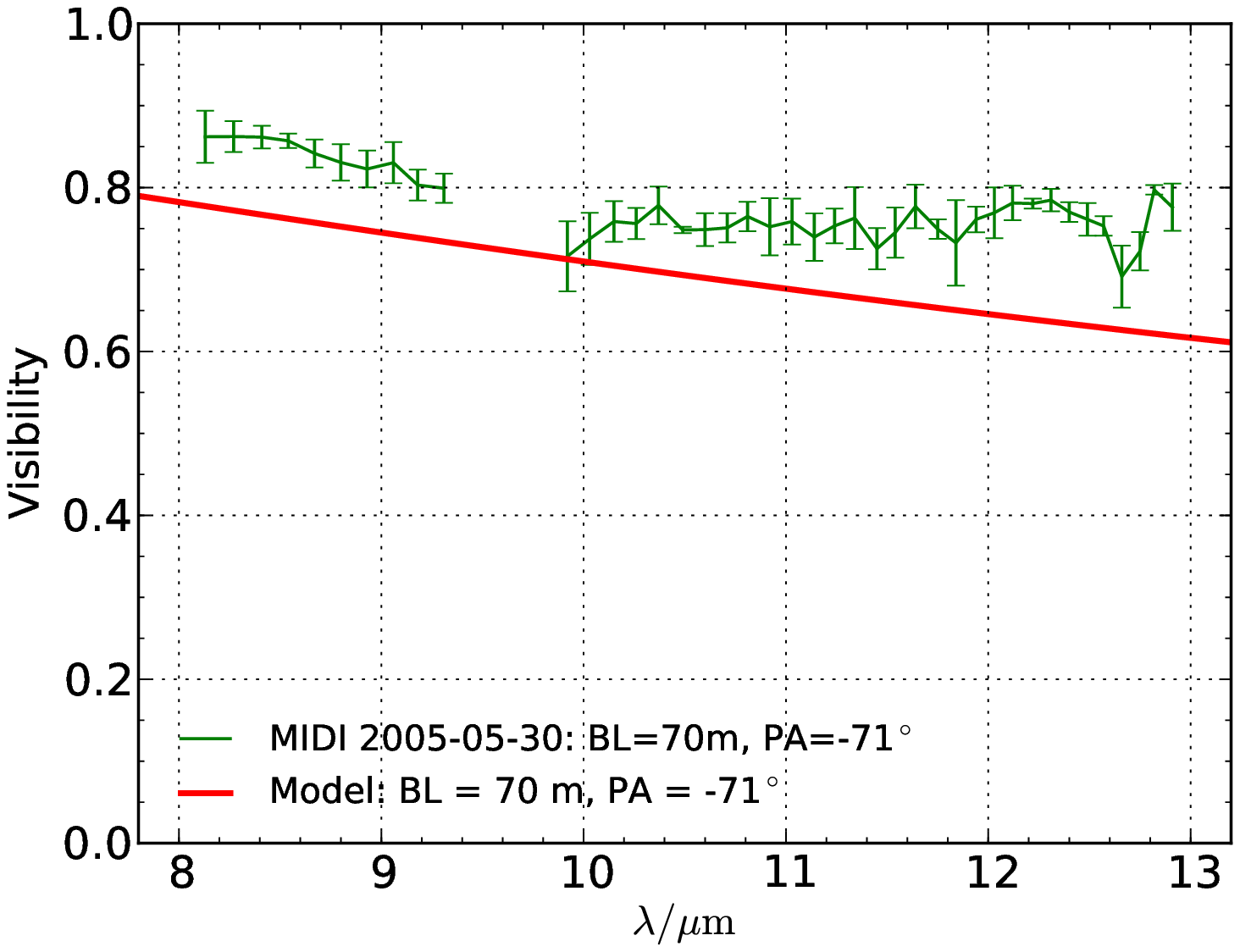}
\begin{minipage}{0.49\textwidth}
  \includegraphics[width=\textwidth]{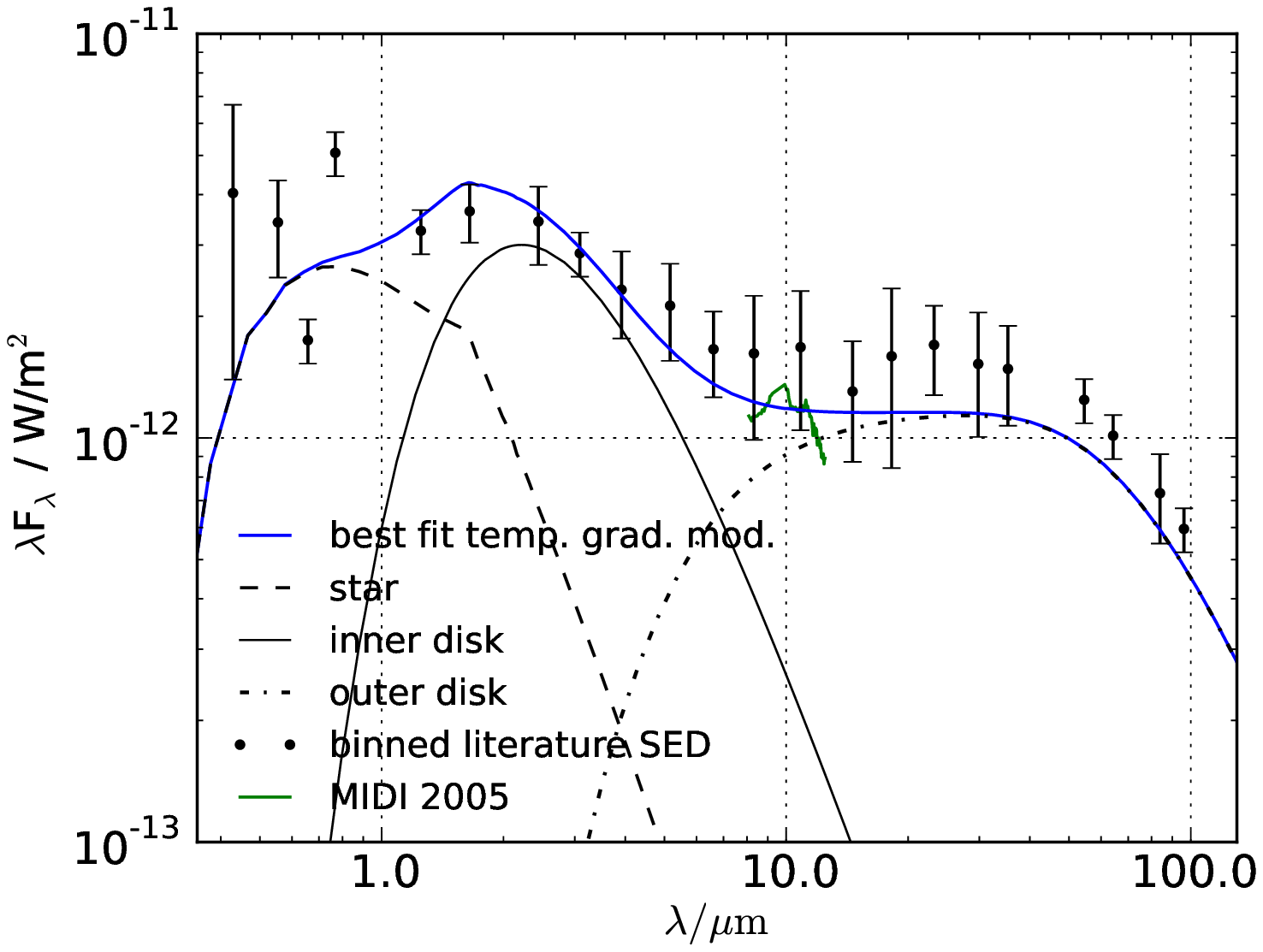}
\end{minipage}
\begin{minipage}{0.5\textwidth}
  \includegraphics[angle=-90,width=\textwidth]{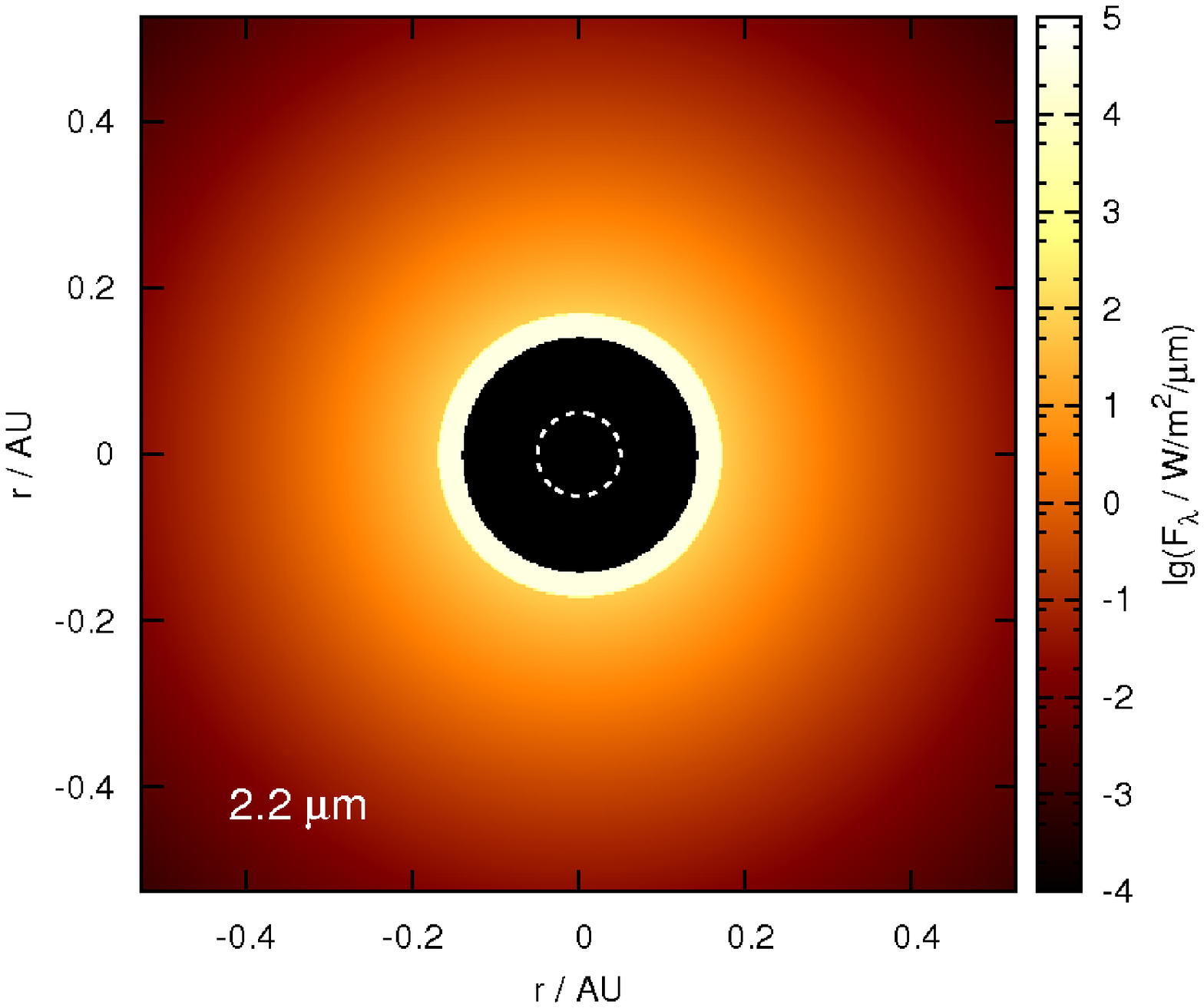}
  \end{minipage}
  \caption{Temperature-gradient model C of S~CrA~N -- {\it Two top
      rows}: NIR visibility. The model that fits visibility and SED
    simultaneously (model C in Table~\ref{tabpar}) is indicated with
    the solid red lines. The six figures show the modeling of the data
    sets I (top) and II (bottom), described in Table~\ref{tabobs2}.
    {\it Third row}: MIDI visibility. The two figures show the
    comparison between our model (red line) and MIR data adopted from
    \citet{2009schwol}. {\it Bottom left:} Spectral energy
    distribution. To model the SED of S~CrA~N, we collected values
    from the literature and binned them (black dots are the dereddened
    SED points, see Appendix~\ref{appsed}). The green line represents
    the MIDI spectrum of \citet{2009schwol}. The SED of model~C is
    indicated with the blue line. It consists of the stellar
    contribution (Kurucz model, black dashed line) plus the
    contribution of two additional ring-like structures (black solid
    and dash-dotted line). {\it Bottom right:} Intensity distribution
    of the temperature-gradient disk of model~C. The dashed white ring
    indicates the expected dust sublimation radius of 0.05~AU
    predicted by the size-luminosity-relation (see
    Sect.~\ref{kapdis}). }
  \label{figtgm2} 
\end{figure*}

Fig.~\ref{figtgm2} shows the intensity distribution of model~C,
together with all S~CrA~N observations. An extended ring is located
between 0.14~AU and 26~AU, whereas the inner ring extends radially
from 0.11~AU to 0.14~AU and has a temperature of 1690~K at the inner
ring edge (see parameters in Table~\ref{tabpar}).

%%%%%%%%%%%%%%%%%%%%%%%%%%%%%%%%%%%%%%%%%%%%%%%%%%%%%%%%%%%%%%%%%%%%%%%%%

%%%%%%%%%%%%%%%%%%%%%%%%%%%%%%%%%%%%%%%%%%%%%%%%%%%%%%%%%%%%%%%%%%%%%%%%%
\section{Discussion} \label{kapdis}
\begin{figure}[b]
\hspace*{-6.5mm}
  \includegraphics[width=0.37\textwidth,angle=-90]{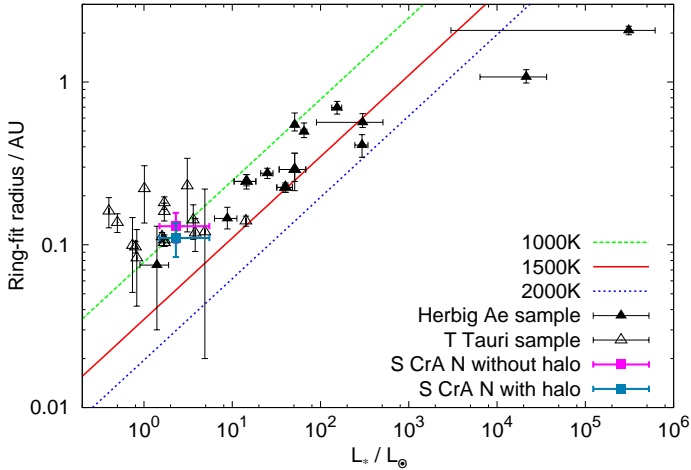}
  \caption{Location of S~CrA~N in the size-luminosity diagram: K-band
    ring-fit radii derived with a geometric ring-star model (pink
    square, $\approx$$0.13$~AU) and a ring-star-halo model
    (blue square, $\approx$$0.11$~AU); see
    Tables~\ref{tabrad} and \ref{tabpro} for the stellar parameters
    used. The lines indicate the predicted theoretical dependence of
    the {\it K}-band radius on the luminosity
    \citep{2002monmil}. Predictions of models including backwarming
    are discussed in Sect.~\ref{kapdis}. The filled black squares are
    a sample of Herbig Ae stars \citep{2005monmil}. The open black
    squares represent a sample of TTS \citep{2008pinmen}.}
  \label{figslr}
\end{figure}

For comparison with other pre-main sequence stars, we plot the
geometric {\it K}-band ring-fit radii, where
$r_\mathrm{ring,in}\approx 0.13$~AU for the geometric model without a
halo and $\approx$$0.11$~AU for the model with a halo;
Sect.~\ref{kapgeo}, of S~CrA~N in the size-luminosity diagram that
shows the {\it K}-band ring-fit radius as a function of the stellar
luminosity $L_*$ (\citealt{2002monmil}, Fig.~\ref{figslr}). The figure
shows a sample of Herbig Ae stars (filled squares,
\citealt{2005monmil}), for which this correlation has been originally
found. Additionally, we plot a sample of TTS (open squares,
\citealt{2008pinmen}).

Figure~\ref{figslr} shows that our measurements of the {\it K}-band
ring-fit radius of S~CrA~N of $\approx$$0.11$--$0.13$~AU, radii
derived with the geometric ring-star and ring-star-halo models, is
approximately 2.4 times larger than the dust sublimation radius of
$\approx$$0.05$~AU predicted for the silicate dust sublimation
temperature of 1500~K and gray dust opacities
\citep{2002monmil}. However, several effects can influence the inner
model radius, such as the chemistry and grain size of the dust
(e.g. \citealt{2002monmil}) or magnetospherical disk truncation
\citep{2007eishil}.  \citet{2009schwol} modeled their mid-infrared
MIDI data and the SED of S~CrA~N using the Monte Carlo code MC3D
\citep{1999wolhen,2008schwol}. They found that an assumed sublimation
radius of 0.05~AU agrees with their observations.

Furthermore, we compared our ring-fit radii of
$\approx$$0.11$--$0.13$~AU derived from the geometric star-disk and
star-disk-halo models and the inner disk radii of our
temperature-gradient models (0.11--0.14~AU) with the predictions of a
model that accounts for both backwarming and accretion luminosity
\citep{2007milmal,2010dulmon}. This model suggests an inner disk
radius of 0.12~AU for a stellar luminosity of 2.3~$L_\odot$ and an
accretion luminosity of 0.6~$L_\odot$ (derived from
\citealt{2003pragre,1998muzhar}), which agrees with our measured inner
radius of $\approx$$0.11$--$0.13$~AU.

%______________________________________________________________

\section{Conclusions} \label{kapcon}
We have observed the T~Tauri star S~CrA~N with AMBER in the {\it H}
and {\it K} bands. We have fitted the geometric star-disk and
star-disk-halo models to our visibility data and derived a radius
$r_\mathrm{ring,in}$ of approximately 0.11--0.13~AU in the {\it K}
band (0.095--0.12~AU in the {\it H} band). We compared the position of
S~CrA~N in the size-luminosity diagram with the position of other YSO,
and found that it is above the line expected for a dust sublimation
temperature of 1500~K and gray dust, but within the region of other
TTS. The radius predicted by this size-luminosity relation is
$\approx$$0.05$~AU, whereas the derived ring-fit radius is
$\approx$0.11--0.13~AU (Table~\ref{tabrad}). However, models including
backwarming \citep{2007milmal,2010dulmon} suggest a larger inner disk
radius of $\approx$$0.12$~AU, which agrees with our
derived ring-fit radiii of $\approx$0.11--0.13~AU.

We tested several temperature-gradient models (one- and two-component
disk models, with or without halo). We found that the near- and mid-IR
visibilities, as well as the SED, can approximately be reproduced by a
temperature-gradient model consisting of a two-component ring-shaped
disk and an unresolved star. The favored temperature-gradient model~C
(see Table~\ref{tabpar}) has a temperature of $\approx$$1700$~K at the
inner disk radius of 0.11~AU. The temperature power-law index $q_1$ of
the inner narrow temperature-gradient disk is approximately 0.2, and
the index $q_2$ of the extended outer disk is approximately 0.5, which
suggests a flared irradiated disk structure
\citep{1997chigol}. However, $q_1$ is not well-constrained because of
the narrow width of the inner component of only $\approx$0.03~AU. The
inner temperature-gradient disk radius of 0.11~AU is similar to the
four geometric ring-fit radii of approximately 0.10~AU to 0.13~AU and
to the prediction of models including backwarming
(0.12~AU). Unfortunately, we cannot place any constraints on the gas
within the inner disk radius of $\approx$0.11~AU since the disk is
only partially resolved (all NIR visibilities are $>0.77$) and the
visibility errors are large.  Interestingly, the inner disk components
of all three best-fit temperature-gradient models consist of a very
narrow, hot ring ($T_\mathrm{in,1}\approx 1500$--$1700$~K; ring width
only $\approx$0.03--0.04~AU) surrounded by a colder ($<$600~K) disk
component with an extension of several AU. This size and temperature
structure is similar to the structure of more sophisticated radiative
transfer models including a hot PUIR (e.g.,
\citealt{2001natpru,2001duldom}). This suggests that the derived
narrow, inner disk component in the temperature-gradient models is
caused by a hot, perhaps PUIR, in S~CrA~N. This rim may explain the
steep temperature jump from 1500 to 600~K.

\begin{acknowledgements}
  J. Vural was supported for this research through a stipend from the
  International Max Planck Research School (IMPRS) for Astronomy and
  Astrophysics at the Universities of Bonn and Cologne
\end{acknowledgements}

%\newpage
\bibliographystyle{aa} 
\bibliography{BIBO}

\begin{appendix}
  
  \section{SED references} \label{appsed}
\begin{figure}[hbt]
    \centering
    \includegraphics[width=0.5\textwidth]{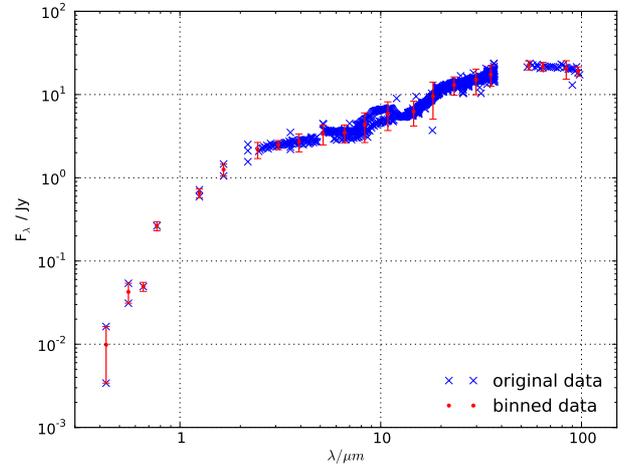}
    \vspace{-2mm}
    \caption{SED of S~CrA~N -- original data according to the sources
      in Table~\ref{tabmes} (blue crosses) and the binned data (red
      dots).}
    \label{figbin}
  \end{figure}

  To simplify the fitting process, we did not use the pile of original
  data (Table~\ref{tabmes}), but instead binned them to form a smaller
  number of data points. Figure~\ref{figbin} shows both the original
  data (blue crosses) and the binned data (red dots). The measurements
  were not recorded contemporaneously (see Table \ref{tabmes}) and
  therefore variability is one of the error sources. Other SED error
  sources are the different aperture sizes and the small number of
  observations in the visible and NIR.

  \begin{table}[h!]
    \caption{Measurements from the literature that were used to
      produce the binned SED of S~CrA~N. A beam width of 10$"$ or
      larger most probably also includes the light from the secondary
      S~CrA~S (separation 1.4$"$).}
    \label{tabmes}
    \begin{tabular}{clr}
       \hline \hline
      Wavelength range &  Reference & Beam \\ \hline
      2.47-11.62 & ISO spectrum & \\
      3.5-170 & ISO & \\
      53.6-106 & Spitzer MIPS & 2.7 x 0.34$'$\\
      5.13-36.9 & Spitzer IRS & 55x81$"$\\
      12-100 & IRAS & 10$"$ \\
      450-1100 & JCMT (upper limit) & 10$"$ \\
      12-100 & Gezari catalog & 10$"$ \\
      0.768 & Denis 2005 & 0.5$"$  \\
      0.429, 0.554 & Tycho catalog & 10$"$  \\
      0.657, 0.429 & USNO & 10$"$  \\
      0.554 & WDS & 0.5$"$  \\
      1.25-3.55 & Prato 2003 & 0.5$"$  \\
      2.18-18.1 & McCabe 2005 & 0.5$"$  \\
      1.25-3.55 & Morel 1978 & 10$"$  \\ \hline
    \end{tabular}
  \end{table}
\end{appendix}

\end{document}